\documentclass[letterpaper,journal]{IEEEtran}
\usepackage[switch]{lineno}
\usepackage{amsmath,amsfonts,amssymb}
\usepackage{algorithmic}
\usepackage{algorithm}
\usepackage{array}
\usepackage[dvipsnames]{xcolor}
\usepackage[caption=false,font=footnotesize,labelfont=sf,textfont=sf]{subfig}
\usepackage{textcomp}
\usepackage{stfloats}
\usepackage{url}
\usepackage{bm}
\usepackage{cite}
\usepackage{hyperref}
\hypersetup{colorlinks=false, linkcolor=blue, citecolor=blue, urlcolor=blue}
\usepackage{verbatim}
\usepackage{graphicx}
\usepackage{siunitx}
\usepackage{float}
\def\BibTeX{{\rm B\kern-.05em{\sc i\kern-.025em b}\kern-.08em
		T\kern-.1667em\lower.7ex\hbox{E}\kern-.125emX}}
\usepackage{balance}

\makeatletter
\renewcommand{\maketag@@@}[1]{\hbox{\m@th\normalsize\normalfont#1}}%
\newcommand{\setAlgoSmall}{%
	\footnotesize 
	\renewcommand{\maketag@@@}[1]{\hbox{\m@th\footnotesize\normalfont##1}}
}
\makeatother

\begin{document}
	\title{\LARGE{Rotatable Antenna Enhanced Multicast Communication System}}
	\author{Weihua Zhu, Beixiong Zheng, \textit{Senior Member, IEEE,} Lipeng Zhu, \textit{Member, IEEE,} Jie~Tang,~\textit{Senior~Member,~IEEE,} and Yong Zeng, \textit{Fellow, IEEE}  \IEEEmembership{}
		\vspace{-0.4cm}	
		\thanks{Weihua Zhu and Beixiong Zheng are with the School of Microelectronics, South China University of Technology, Guangzhou 511442, China (e-mails: miweihuazhu@mail.scut.edu.cn; bxzheng@scut.edu.cn).}
		\thanks{Lipeng Zhu is with the State Key Laboratory of CNS/ATM and the School of Interdisciplinary Science, Beijing Institute of Technology, Beijing 100081, China (E-mail: zhulp@bit.edu.cn).}
		\thanks{Jie Tang is with the School of Electronic and Information Engineering, South China University of Technology, Guangzhou 510640, China (e-mail:eejtang@scut.edu.cn).}
		\thanks{Yong Zeng is  with the National Mobile Communications Research Laboratory, Southeast University, Nanjing 210096, China, and also with the Purple Mountain Laboratories, Nanjing 211111, China (e-mail: yong\_zeng@seu.edu.cn).}}
	
	
	\maketitle
	
	\begin{abstract}
		Rotatable antenna~(RA) provides additional spatial degrees of freedom~(DoFs) for communication systems by enabling per-antenna dynamic boresight adjustment, which is attractive for fairness-oriented multicast transmission. This letter investigates an RA-enhanced downlink multi-group multicast system. Specifically, we aim to maximize the minimum signal-to-interference-plus-noise ratio~(SINR) among all users by jointly optimizing the multicast beamforming vectors and the RA boresight directions under transmit power and rotation constraints. To solve this non-convex problem, we first reformulate the max–min SINR objective via quadratic transform. Then, we develop an alternating optimization~(AO) algorithm that iteratively updates the multicast beamforming and RA boresight directions. The beamforming vectors are obtained from a convex subproblem, while the boresight directions are refined using a successive convex approximation~(SCA) procedure. Simulation results verify that the proposed RA-based scheme substantially enhances the fairness performance compared with fixed antenna-based and random-orientation benchmarks.
	\end{abstract}
	
	
	\vspace{-0.3cm}
	\section{Introduction}
	\vspace{-0cm}
	With the rapid proliferation of wireless devices and high-rate multimedia services, many applications require delivering identical content to groups of users (e.g., emergency broadcasting, video streaming, and system updates)~\cite{zaherCellFreeBeamformingDesign2025}. Multi-group multicast serves multiple groups with distinct messages in parallel, thus improving spectral efficiency and reducing signaling overhead compared with repeated unicast transmissions. However, each group’s achievable rate is bottlenecked by its worst-channel user and can be further degraded by inter-group interference, making fairness guarantees particularly challenging in heterogeneous three-dimensional~(3D) user distributions.
	
	Recently, flexible antenna architectures have been proposed to exploit additional spatial degrees of freedom~(DoFs) by reconfiguring the radiation characteristics of the array in space. Representative technologies include movable antenna (MA)~\cite{zhuModelingPerformanceAnalysis2024a,zengWirelessCommunicationforLowAltitudeEconomy}, six-dimensional movable antenna (6DMA)~\cite{shaoTutorialSixDimensionalMovable2025}, and fluid antenna system (FAS)~\cite{newFluidAntennaSystems2025}, which dynamically adapt antenna positions and/or orientations to the channel environment. Within this family, the rotatable antenna~(RA) architecture is particularly attractive. Its key distinction from conventional mechanical counterparts~(e.g., mechanical scanning radar) is the ability to perform element-scale reconstruction of the boresight direction. By enabling per-antenna 3D boresight rotation with fixed positions, RA avoids complex position-orientation coupling and allows for compact implementations. Given these benefits, initial research has focused on channel modeling and channel estimation for RA-enabled systems~\cite{wuModelingOptimizationRotatable2025,zhengRotatableAntennaEnabled2025,xiongEfficientChannelEstimation2025}. Then, the potential of RA across diverse scenarios 
	was investigated~\cite{daiRotatableAntennaEnabledSecure2025,zhengRotatableAntennaEnabled2025a,xiongIntelligentRotatableAntenna2025}. For instance, RA has been utilized in physical layer security (PLS)~\cite{daiRotatableAntennaEnabledSecure2025}, integrated sensing and communication (ISAC)~\cite{zhengRotatableAntennaEnabled2025a}, and integrated sensing, communication, and computation (ISCC)~\cite{xiongIntelligentRotatableAntenna2025}, showing performance gains not only in traditional communications but also in new emerging applications. Notably, through flexible 3D boresight adjustment, RA can balance the array directional gain over multiple users~\cite{zhengRotatableAntennaEnabled2025}, which is particularly advantageous to tackle the worst-user bottleneck issue in multicast systems. Nevertheless, to the best of our knowledge, the existing RA-related research has predominantly concentrated on unicast transmissions, while RA-enhanced multi-group multicast transmission has not yet been investigated.
	\begin{figure}[!t]
		\centering
		\includegraphics[width=2.3in]{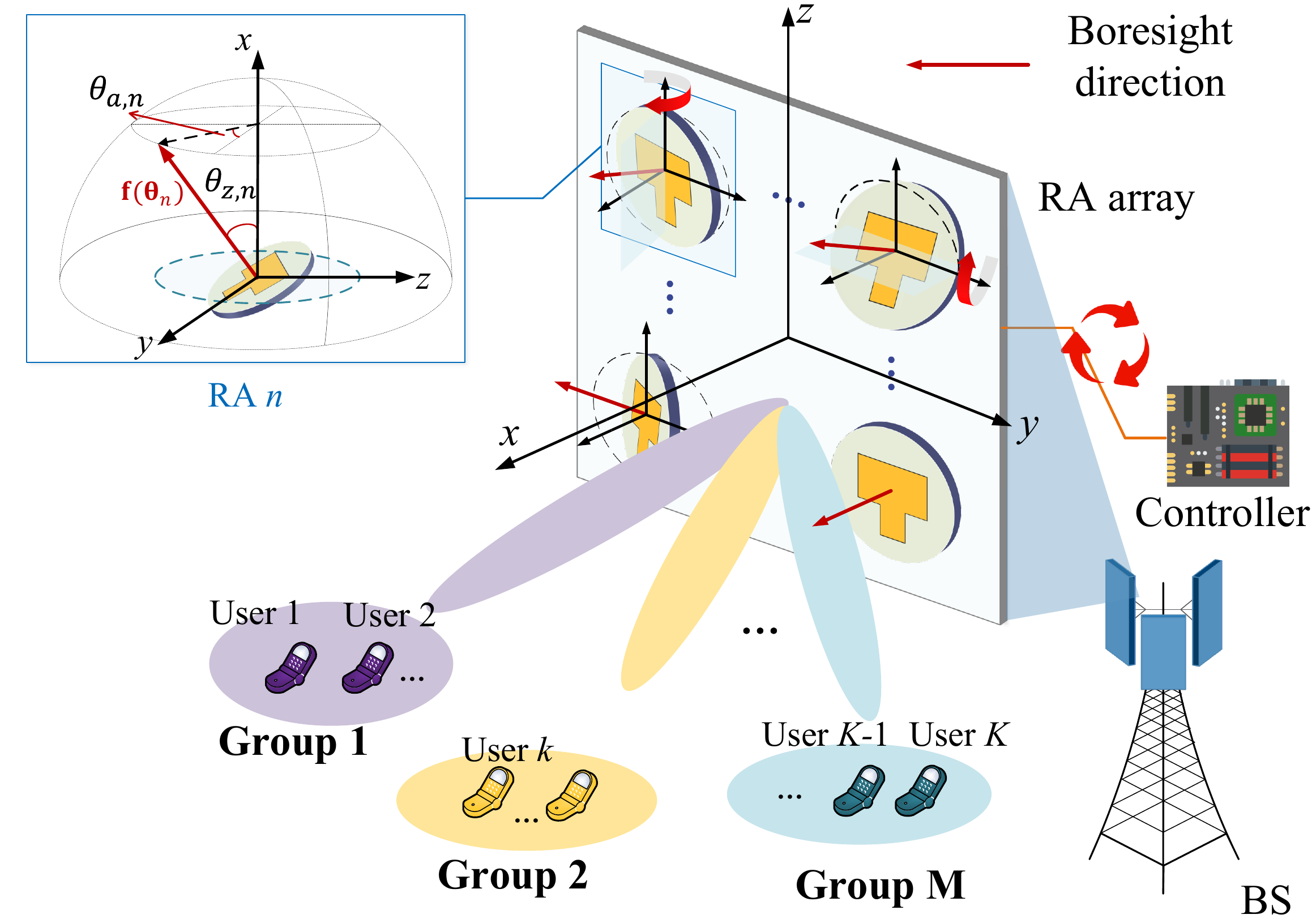}
		\caption{Illustration of system model, $\theta_{z,n}$ and $\theta_{a,n}$.}
		\label{fig:system model}
	\end{figure}
	
	Motivated by the above, in this letter we investigate an RA-enhanced downlink multi-group multicast system, where a base station~(BS) equipped with RAs serves multiple multicast groups, as illustrated in Fig.~\ref{fig:system model}. Our objective is to maximize the minimum signal-to-interference-plus-noise ratio~(SINR) among all users by jointly optimizing the multicast beamforming vectors and the RA boresight directions under the transmit power and rotation constraints. To tackle the resultant non-convex max-min SINR problem with coupled beamforming and boresight directions, we employ a quadratic transform and design an alternating optimization~(AO) algorithm. This algorithm iteratively updates the beamforming via a convex quadratically constrained program~(QCP) and refines the RA boresight directions via a successive convex approximation~(SCA) procedure. Simulation results demonstrate that the proposed RA-based scheme significantly improves the max-min SINR over fixed directional antenna-based, random-orientation, and isotropic antenna-based benchmark schemes.

	\vspace{-0.2cm}
	\section{System Model and Problem Formulation}
	\subsection{System Model}
	\vspace{-0.1cm}
	As illustrated in Fig.~\ref{fig:system model}, we consider a downlink multi-group multicast system, where a BS serves $K$ single-antenna users. Specifically, the BS is equipped with a uniform planar array (UPA) consisting of $N$ RAs, and users are partitioned into $M$ disjoint groups. Moreover, we let $\mathcal{K} \triangleq \{1,\ldots,K\}$ and $\mathcal{M} \triangleq \{1,\ldots,M\}$ denote the user and group index sets, respectively. For each $m \in \mathcal{M}$, let $\mathcal{G}_m \subseteq \mathcal{K}$ denote the set of users in group $m$. Since each user belongs to only one group, we have $\mathcal{G}_m \cap \mathcal{G}_j = \emptyset$, $\forall\, m,j \in \mathcal{M}$ with $m \neq j$, and $\bigcup_{m=1}^{M} \mathcal{G}_m = \mathcal{K}$.
	
	We assume that the UPA at the BS lies on the $y$-$z$ plane of a 3D Cartesian coordinate system, with its geometric center located at the origin. The boresight direction of RA\:$n$ for $n\in\mathcal{N}\triangleq\{1,\ldots,N\}$ can be parameterized by $\boldsymbol{\theta}_{n}\triangleq[\theta_{z,n},\,\theta_{a,n}]^{{T}}$, where $\theta_{z,n}$ denotes the zenith angle of RA$\;n$, i.e., the angle between its boresight and the $+x$-axis, and $\theta_{a,n}$ denotes the azimuth angle, i.e., the angle between the projection of the boresight direction onto the $y$-$z$ plane and the $+y$-axis. Thus, the 3D boresight direction of RA~$n$ can be characterized by a unit pointing vector $\mathbf{f}(\boldsymbol{\theta}_{n})$. Based on the standard spherical-to-Cartesian mapping, this vector is given by
	
	\vspace{-0.5cm}
		\begin{equation}
			\label{BoreSightVec}
			\mathbf{f}(\boldsymbol{\theta}_{n})=\big[\sin\theta_{z,n}\cos\theta_{z,n},\cos\theta_{a,n}, \sin\theta_{z,n}\sin\theta_{a,n}\big]^{{T}},
		\end{equation}
	which satisfies $\lVert\mathbf{f}(\boldsymbol{\theta}_{n})\rVert = 1$ by definition, where $\lVert \cdot \rVert$ denotes the Euclidean norm. According to \cite{daiRotatableAntennaEnabledSecure2025,zhengRotatableAntennaEnabled2025}, the directional gain of each RA element can be modeled by a generic cosine-power pattern as follows:
	
	\vspace{-0.2cm}
	\begin{small}
		\begin{equation}
			\label{GenericGainPattern}
			\begin{array}{c}
				G(\epsilon)=
				\begin{cases}
					G_0\cos^{2p}(\epsilon),& 0\leq \epsilon < \tfrac{\pi}{2},\\
					0,& \tfrac{\pi}{2}\leq \epsilon \leq \pi,
				\end{cases}
			\end{array}
		\end{equation}
	\end{small}\normalsize
	
	\vspace{-0.1cm}
	\noindent
	where $\epsilon$ is the incident angle with respect to~(w.r.t.) the RA boresight, $p> 0$ is the directivity factor that characterizes the beamwidth of the antenna's main lobe, and $G_0=2(2p+1)$ follows the power conservation \cite{zhengRotatableAntennaEnabled2025a}.
	
	The incident angle $\epsilon_{k,n}$ for the line-of-sight~(LoS) path of user $k$\:(located at $\mathbf{q}_k \in \mathbb{R}^{3\times1}$) w.r.t. RA\:$n$\:(located at $\mathbf{p}_n \in \mathbb{R}^{3\times1}$) is defined by $\cos(\epsilon_{k,n})=\mathbf{f}(\boldsymbol{\theta}_{n})^{{T}}\vec{\mathbf{u}}_{k,n}$, with the unit vector $\vec{\mathbf{u}}_{k,n}\triangleq\frac{\mathbf{q}_{k}-\mathbf{p}_{n}}{\lVert \mathbf{q}_{k}-\mathbf{p}_{n}\rVert}$. We consider the near-field LoS-dominant channel model. Specifically, the channel from RA\:$n$ to user \:$k$ is given by
	 
	\vspace{-0.45cm}
	\begin{equation}
		\label{ChanRA_n_User_k}
		h_{k,n}(\mathbf{f}_{n})=\sqrt{g_{k,n}(\mathbf{f}_{n})}e^{-j\frac{2\pi }{\lambda}d_{k,n}}=\tilde{\beta}_{k,n}(\mathbf{f}_{n}^{T}\vec{\mathbf{u}}_{k,n})^{p},
	\end{equation}
	
	\vspace{-0.15cm}
	\noindent
	where $\mathbf{f}_{n}\triangleq\mathbf{f}(\boldsymbol{\theta}_{n})$ for notational simplicity, and $g_{k,n}(\mathbf{f}_{n})=\frac{S}{4\pi d_{k,n}^2}G_0\cos^{2p}(\epsilon_{k,n})$ denotes channel gain between RA\:$n$ and user\:$k$, with $S$ being the physical size of each RA, and $\lambda$ is the carrier wavelength. In addition, $\tilde{\beta}_{k,n}=\sqrt{\frac{S}{4\pi d_{k,n}^{2}}G_0}e^{-j\frac{2\pi}{\lambda}d_{k,n}}$. 
	
	Stacking the $N$ per-antenna coefficients for user $k$ yields the following channel vector
	
	\vspace{-0.3cm}
	\begin{equation}
		\label{ChannelVector}
		\mathbf{h}_{k}(\mathbf{F})=\big[h_{k,1}(\mathbf{f}_{1}),h_{k,2}(\mathbf{f}_{2}),\ldots,h_{k,N}(\mathbf{f}_{N})\big]^{{T}},
	\end{equation}
	
	\vspace{-0.1cm}
	\noindent
	where $\mathbf{F}\triangleq [\mathbf{f}_{1},\mathbf{f}_{2},\ldots,\mathbf{f}_{N}]\in \mathbb{R}^{3\times N}$. For the multi-group multicast transmission, let $\mathbf{w}_{m}\in \mathbb{C}^{N\times 1}$ denote the linear beamforming vector for group $m\in\mathcal{M}$, and let $s_{m}$ be its data symbol with $\mathbb{E}\{\lvert s_{m} \rvert^{2}\}=1$. The received signal at user $k\in \mathcal{G}_{m}$ is then given by
	
	\vspace{-0.6cm}
	\begin{equation}
		\label{ReceivedSignalUserk}
		y_{k} = \mathbf{h}_{k}(\mathbf{F})^{T}\mathbf{w}_{m}s_{m}
			+ \sum_{j \neq m}^{M}
			\mathbf{h}_{k}(\mathbf{F})^{T}\mathbf{w}_{j}s_{j}
			+ n_{k},
	\end{equation}
	
	\vspace{-0.2cm}
	\noindent
	where $n_{k}\sim\mathcal{CN}(0,\sigma_{k}^{2})$ denotes the additive white Gaussian noise (AWGN) at user $k$. In the multi-group multicast setting, each group shares a distinct common message. As a result, the receive SINR of user $k\in\mathcal{G}_{m},\:m\in\mathcal{M}$, denoted by $\gamma_k(\mathbf{W},\mathbf{F})$ with $\mathbf{W} \triangleq [\mathbf {w}_1,\mathbf{w}_2,\ldots,\mathbf{w}_M]\in\mathbb C^{N\times M}$, is given by
	
	\vspace{-0.2cm}
	\begin{equation}
		\label{ReceivedSINR_k}
		\gamma_{k}\left(\mathbf{W} ,\mathbf{F}\right)=\frac{\lvert\mathbf{h}_{k}(\mathbf{F})^{T}\mathbf{w}_{m}\rvert^{2}}{\sum_{j\neq m}^{M}\lvert\mathbf{h}_{k}(\mathbf{F})^{T}\mathbf{w}_{j} \rvert^{2}+\sigma_{k}^{2}}.
	\end{equation}

	\subsection{Problem Formulation}
	
	Under the multicast setting, each group shares a single common stream, and successful decoding requires every user in that group to attain a sufficient signal quality. A natural fairness criterion is to maximize the minimum user SINR across all groups. Specifically, the optimization problem can be expressed as
	
	\vspace{-0.5cm}
		\begin{subequations}
			\label{OptProOri:OptProblem_1}
			\begin{align}
				\text{(P1):}\hspace{-10mm}&& \underset{\mathbf{W} ,\mathbf{F}}{\text{max}} 
				\quad& \underset{k\in\mathcal{K}}{\text{min}} \quad \gamma_{k}(\mathbf{W},\mathbf{F}) \label{OptProOri:obj}\\
				&& \text{s.t.} \quad& \mathbf{f}_{n}^{T}\mathbf{e}_{x}\geq\cos(\theta_{\max}),\;\forall n \in \mathcal{N}, \label{OptProOri:constr1} \\
				&& \phantom{\text{s.t.} \quad}& \|\mathbf{f}_{n}\|=1,\;\forall n \in \mathcal{N}, \label{OptProOri:constr2}\\
				&& \phantom{\text{s.t.} \quad}& \sum_{m=1}^{M}\lVert\mathbf{w}_{m}\rVert^{2} \leq P_t, \label{OptProOri:constr3}
			\end{align}
		\end{subequations}
	\normalsize
	
	\vspace{-0.3cm}
	\noindent	
	where $\mathbf{e}_{x}=\left[1,0,0\right]^{T}$ represents the unit pointing vector in the $+x$ axis direction. Constraint \eqref{OptProOri:constr1} confines the zenith deflection angles $\{\theta_{z,n}\}_{n\in\mathcal{N}}$ to the interval $[0,\theta_{\text{max}}]$, where $\theta_{\text{max}}$ denotes the maximum allowable zenith angle for each RA.\footnote{The maximum zenith angle $\theta_{\text{max}}$ is imposed by the hardware limitations of the RA elements and is also used to mitigate mutual coupling between adjacent elements, as discussed in  \cite{zhengRotatableAntennaEnabled2025}.} Constraint \eqref{OptProOri:constr3} limits the maximum transmit power of the BS to $P_t$. Note that problem (P1) is inherently non-convex, since the SINR expressions are fractional quadratic functions of the beamforming vectors and exhibit a nonlinear dependence on the pointing vectors through the RA's gain pattern. Moreover, the intrinsic coupling between the beamforming and pointing vectors precludes joint convexity in $(\mathbf{W} ,\mathbf{F})$, thereby motivating the AO framework developed in the next section.
	
	\vspace{-0.3cm}
	\section{Proposed Algorithm}
	In this section, we first convert problem (P1) into a more tractable equivalent form via quadratic transform. Then, based on the converted equivalent problem, we adopt an AO scheme to handle the strong coupling between the beamforming matrix and pointing vectors by updating them alternately. Specifically, we update the multicast beamforming matrix $\mathbf{W}$ by solving a QCP problem under the power constraint and refine the pointing vectors via SCA technique. 
	
	\vspace{-0.35cm}
	\subsection{Conversion of Original Problem}
	To tackle the fractional structure in \eqref{OptProOri:obj}, we apply quadratic transform in  \cite{shenFractionalProgrammingCommunication2018a} and reformulate problem (P1) as
	
	\vspace{-0.5cm}
		\begin{subequations} 
			\label{OriProQT:QTforOriprolem}
			\begin{align}
				\text{(P2):}\hspace{-2mm}\quad\underset{t, \mathcal{Z},\mathbf{W},\mathbf{F}}{\text{max}}\quad&
				t \label{OriProQT:Obj} \\
				\text{s.t.} \quad&	
				\tilde{\gamma}_{k}(z_{k},\mathbf{W},\mathbf{F})\geq t,\;\forall k \in \mathcal{K},\label{OriProQT:constr1}\\
				\phantom{\text{s.t.}\quad}&\eqref{OptProOri:constr1},\eqref{OptProOri:constr2},\eqref{OptProOri:constr3}, \label{OriProQT:constr2}
			\end{align}
		\end{subequations}
	where $t$ is an auxiliary variable, and $\mathcal{Z}\triangleq\left\lbrace z_{k}\right\rbrace_{k\in\mathcal{K}}$ denotes the set of auxiliary complex variables introduced by the quadratic transform. $\tilde{\gamma}_{k}(z_k,\mathbf{W},\mathbf{F})$ (for $k\in\mathcal{G}_{m}, m\in\mathcal{M}$) is a surrogate function, which is given by
	
	\vspace{-0.3cm}
	\begin{small}
	\begin{equation}
		\begin{aligned}
			\tilde{\gamma}_{k}(z_{k},\mathbf{W},\mathbf{F})=&\:2\operatorname{Re}\left\lbrace z_{k}^{*}\mathbf{h}_{k}(\mathbf{F})^{T}\mathbf{w}_{m}\right\rbrace\\
				-&\lvert z_{k}\rvert^{2}\left(\sum_{j\neq m}^{M}\lvert\mathbf{h}_{k}(\mathbf{F})^{T}\mathbf{w}_{j} \rvert^{2}+\sigma_{k}^{2}\right),
		\end{aligned}	
	\end{equation}
	\end{small}\normalsize
	where $z_{k}^{*}$ is the conjugate of $z_{k}$. For any given $\mathbf{W}$ and $\mathbf{F}$, the optimal $z_{k}$ has the following closed-form expression:
	
	\vspace{-0.2cm}
		\begin{equation}
			\label{Opt_eta}
			z_{k}^{\star}(\mathbf{W},\mathbf{F}) = \frac{\mathbf{h}_{k}(\mathbf{F})^{T}\mathbf{w}_{m}}{\sum_{j\neq m}^{M}\lvert\mathbf{h}_{k}(\mathbf{F})^{T}\mathbf{w}_{j} \rvert^{2}+\sigma_{k}^{2}}.
		\end{equation}
	The reformulation in \eqref{OriProQT:QTforOriprolem} thus converts the original fractional objective into a more tractable form. Note that (P2) is equivalent to (P1), since (P2) is constructed via epigraph reformulation from (P1), and $\tilde{\gamma}_{k}(z_{k}^{\star},\mathbf{W},\mathbf{F})=\gamma_{k}\left(\mathbf{W} ,\mathbf{F}\right)$. Furthermore, to tackle the coupling between multicast beamforming and pointing vectors, an AO framework is adopted.
	
	\vspace{-0.4cm}
	\subsection{Optimization for Beamforming Vectors}
	Given all the RA pointing vectors, problem (P2) reduces to a multigroup multicast beamforming problem:
	
	\vspace{-0.5cm}
		\begin{subequations}
			\label{SubPro1:Beamforming}
			\begin{align}
				\text{(P3):}~\underset{t, \mathcal{Z},\mathbf{W}}{\text{max}}\quad&
				t \label{SubPro1:Obj} \\
				\text{s.t.} \quad&	
				2\mathrm{Re}\left\lbrace{z}_{k}^{*}\mathbf{h}_{k}(\mathbf{F})^{T}\mathbf{w}_{m}\right\rbrace\nonumber\\
				\phantom{\text{s.t.}\quad}&-\lvert z_{k}\rvert^{2}\left(\sum_{j \neq m}^{M}\mathbf{w}_{j}^{H}{\mathbf H}_{k}(\mathbf{F})\mathbf{w}_{j}+\sigma_{k}^{2}\right)\geq t,\nonumber\\
				\phantom{\text{s.t.}\quad}&\forall k \in \mathcal{G}_{m},\; \forall m \in \mathcal{M},\label{SubPro1:constr1}\\
				\phantom{\text{s.t.}\quad}&\eqref{OptProOri:constr3}, \label{SubPro1:constr2}
			\end{align}
		\end{subequations}
	\normalsize
	
	\vspace{-0.2cm}
	\noindent
	where $\mathbf{H}_{k}(\mathbf{F})\triangleq\mathbf{h}_{k}(\mathbf{F})^{*}\mathbf{h}_{k}(\mathbf{F})^{T}$, and $\mathcal{Z}$ is updated in closed form by \eqref{Opt_eta}. Given $\left(\mathbf{F},\mathcal{Z}\right)$, each constraint in \eqref{SubPro1:constr1} is a concave quadratic function of $\mathbf{W}$ lower-bounded by $t$; together with the convex power constraint \eqref{OptProOri:constr3}, making this subproblem a convex QCP solvable by CVX~\cite{grant2014cvx}.
	
	\vspace{-0.3cm}
	\subsection{Optimization for Pointing Vectors}
	For any given multicast beamforming matrix $\mathbf{W} $, problem \text{(P2)} can be reformulated as
	
	\vspace{-0.5cm}
		\begin{subequations} 
			\label{SubPro2:DeflctionAngOpt}
			\begin{align}
				\text{(P4):}~\underset{t,\mathcal{Z},\mathbf{F}}{\text{max}}\quad&
				t \label{SubPro2:Obj} \\
				\text{s.t.} \quad&	
				\tilde{\gamma}_{k}({z}_{k},\mathbf{W},\mathbf{F})\geq t,\;\forall k \in \mathcal{K},\label{SubPro2:constr1}\\
				\phantom{\text{s.t.}\quad}&\mathbf{f}_{n}^{T}\mathbf{e}_{x}\geq\cos(\theta_{\max}),\;\forall n \in \mathcal{N},\label{SubPro2:constr2}\\
				\phantom{\text{s.t.}\quad}&\|\mathbf{f}_{n}\|=1,\;\forall n \in \mathcal{N}.\label{SubPro2:constr3}
			\end{align}
		\end{subequations}
	\normalsize
	Since constraint \eqref{SubPro2:constr1} is non-convex, we employ the SCA technique to approximate it with a convex constraint. This allows us to obtain a local optimum to problem \text{(P4)} iteratively. Specifically, for any given $\{z_{k}\}_{k\in\mathcal{K}}$, we rewrite constraint \eqref{SubPro2:constr1} as
	
	\vspace{-0.6cm}
	
		\begin{equation}
			\label{eq:SigQTSINR}
			\begin{aligned}
				\underbrace{2\:\mathrm{Re}\left\lbrace {z}_{k}^{*}x_{k,m}(\mathbf{F})\right\rbrace}_{\text{Desired term }u_{k}(\mathbf{F})}&-\underbrace{\lvert{z}_{k}\rvert^{2}\sum_{j\neq m}^{M}\lvert x_{k,j}(\mathbf{F})\rvert^{2}}_{\text{Interference term}\:q_{k}(\mathbf{F})}\\
				&\geq t+\lvert {z}_{k}\rvert^{2}\sigma_{k}^{2},
				\:\forall k\in\mathcal{G}_{m}, m\in\mathcal{M}.
			\end{aligned}
		\end{equation}
	\normalsize
	where $x_{k,m}(\mathbf {F})=\sum_{n=1}^{N}\tilde{\beta}_{k,n}w_{m,n}(\mathbf{f}_n^{T}\vec{\mathbf{u}}_{k,n})^{p}$, and $w_{m,n}$ is the $n$-th element of $\mathbf{w}_{m}$.  We define
	
	\vspace{-0.6cm}
		\begin{align}
			\abovedisplayshortskip=0pt
			\belowdisplayshortskip=0pt
			\abovedisplayskip=0pt
			\belowdisplayskip=0pt
			u_{k}(\mathbf{F})&\triangleq 2\,\mathrm{Re}\left\lbrace{z}_{k}^{*}x_{k,m}(\mathbf{F})\right\rbrace=\sum_{n=1}^{N}c_{k,n}g_{k,n}(\mathbf{f}_{n}),\label{eq:DesTerm}\\[-10pt]
			q_{k}(\mathbf{F}) &\triangleq \lvert{z}_{k}\rvert^{2}\sum_{j\neq m}^{M}\lvert x_{k,j}(\mathbf{F})\rvert^{2}=\lvert{z}_{k}\rvert^{2}\sum_{j\neq m}^{M}a_{k,j}(\mathbf{F}),\label{eq:IntTerm}
		\end{align}
	
	\vspace{-0.2cm}
	\noindent
	where $c_{k,n}\triangleq2\,\mathrm{Re}\{z_k^{*}\tilde{\beta}_{k,n}w_{m,n}\},\forall k\in\mathcal{G}_{m},m\in\mathcal{M}$. Moreover, $a_{k,j}(\mathbf{F})=\lvert x_{k,j}(\mathbf{F})\rvert^{2}$ and $g_{k,n}(\mathbf{f}_{n})=(\mathbf{f}_{n}^{T}\vec{\mathbf{u}}_{k,n})^{p}$. 
	To construct a concave lower bound of the left-hand side of \eqref{eq:SigQTSINR}, we formulate corresponding bounds of its components. This is achieved by deriving a concave lower bound of the desired term and a convex upper bound of the interference term at  $\mathbf{F}^{(i)}=\left[\mathbf{f}_{1}^{(i)},\mathbf{f}_{2}^{(i)},\ldots,\mathbf{f}_{N}^{(i)}\right]$, which denotes $\mathbf{F}$  obtained in the $i$-th iteration. Without loss of generality, the update procedure for the $(i+1)$-th iteration is then detailed.
	
	\textit{1) Construct the Lower Bound of $u_{k}$:} The function $g_{k,n}(\mathbf{f}_{n})$ is a power function whose convexity is contingent on $p$. Furthermore, the sign of $c_{k,n}$ varies across iterations, which complicates the convexity analysis of $u_k(\mathbf{F})$. To address this, we adapt the approach in \cite[Lemma 12]{sunMajorizationMinimizationAlgorithmsSignal2017a} and employ a second-order Taylor expansion at $\mathbf{F}^{(i)}$ to derive a local approximation for $u_{k}(\mathbf{F})$:

	\vspace{-0.5cm}
		\begin{align}
			\label{ieq:lb_uk}
			u_{k}(\mathbf{F})&\geq u_{k}(\mathbf{F}^{(i)})+\sum_{n=1}^{N}\left(\nabla_{\mathbf{f}_{n}^{(i)}}u_{k}(\mathbf{F}^{(i)})\right)^{T}\left(\mathbf{f}_{n}-\mathbf{f}_{n}^{(i)}\right)\nonumber\\
			&-\frac{L_{k}^{\mathrm{S}}}{2}\|\mathbf{F}-\mathbf{F}^{(i)}\|_{F}^{2}\triangleq \underline{u}_{k}^{(i+1)}(\mathbf{F}),
		\end{align}
	
	\vspace{-0.2cm}
	\noindent
	where $L_{k}^{\mathrm{S}}>0$ is a Lipschitz constant to ensure $\underline{u}_{k}(\mathbf{F}^{(i)})$ as a conservative lower bound. The construction of $L_{k}^{\mathrm{S}}$ is given in the Appendix. The gradient
	of $u_{k}(\mathbf{F})$ w.r.t. $\mathbf{f}_{n}$ is $\nabla_{\mathbf{f}_{n}}u_{k}(\mathbf{F})=c_{k,n}\nabla g_{k,n}(\mathbf{f}_{n})$.  $\nabla g_{k,n}(\mathbf{f}_{n})\in\mathbb{R}^{3\times1}$ is the gradient of $g_{k,n}(\mathbf{f}_{n})$ w.r.t. $\mathbf{f}_{n}$, which is given by
	
	\vspace{-0.5cm}
	\begin{align}
		\nabla g_{k,n}(\mathbf{f}_{n})&=p(\mathbf{f}_{n}^{T}\vec{\mathbf{u}}_{k,n})^{p-1}\vec{\mathbf{u}}_{k,n}.
	\end{align}	
	
	\textit{2) Construct the Upper Bound of $q_{k}$:}
	We first derive the convex upper bound of $a_{k,j}(\mathbf{F})$. Similar to \eqref{ieq:lb_uk}, with the given local point $\mathbf{F}^{(i)}$, according to \cite{sunMajorizationMinimizationAlgorithmsSignal2017a}, we have
	
	\vspace{-0.5cm}
		\begin{align}
			\label{ieq:ub_akj}
			a_{k,j}(\mathbf{F})&\leq a_{k,j}(\mathbf{F}^{(i)})+\sum_{n=1}^{N}\left(\nabla_{\mathbf{f}_{n}^{(i)}}a_{k,j}(\mathbf{F}^{(i)})\right)^{T}\left(\mathbf{f}_{n}-\mathbf{f}_{n}^{(i)}\right)\nonumber\\[-2pt]
			&+\frac{L_{k,j}^{\mathrm{I}}}{2}\|\mathbf{F}-\mathbf{F}^{(i)}\|_{F}^{2}\triangleq \hat{a}_{k,j}^{(i+1)}(\mathbf{F}),
		\end{align}
	
	\vspace{-0.2cm}
	\noindent
	where $L_{k,j}^{\mathrm{I}}>0$ is another Lipschitz constant given in the Appendix, and $\nabla_{\mathbf{f}_{n}}a_{k,j}(\mathbf{F})\in\mathbb{R}^{3\times1}$ denotes the gradient of $a_{k,j}(\mathbf{F})$ w.r.t. $\mathbf{f}_{n}$, which is derived as
	
	\vspace{-0.3cm}
	\begin{align}	
		\nabla_{\mathbf{f}_{n}}a_{k,j}(\mathbf{F})=2\mathrm{Re}\left\lbrace x_{k,j}(\mathbf{F})^{*}\tilde{\beta}_{k,n}w_{j,n} \right\rbrace	\nabla g_{k,n}(\mathbf{f}_{n}).
	\end{align}
	Therefore, a convex upper bound of $q_{k}(\mathbf{F})$ with the given point $\mathbf{F}^{(i)}$ can be simply expressed as
	
	\vspace{-0.5cm}
		\begin{equation}
			\label{ieq:ub_qk}
			\hat{q}_{k}^{(i+1)}(\mathbf{F})\triangleq\left|{z}_{k}\right|^{2}\sum_{j\neq m}^{M}\hat{a}_{k,j}^{(i+1)}(\mathbf{F}), \forall k\in\mathcal{G}_{m}, m\in\mathcal{M}.
		\end{equation}
		
	\vspace{-0.2cm}
	\textit{3) Convex Problem Reformulation:} Combining \eqref{ieq:lb_uk} with \eqref{ieq:ub_qk} yields a concave surrogate function
	
	\vspace{-0.4cm}
	\begin{equation}
		\label{eq:lb_surrrogate}
		\underline{\phi}_{k}^{(i+1)}(\mathbf{F})=\underline{u}_{k}^{(i+1)}(\mathbf{F})-\hat{q}_{k}^{(i+1)}(\mathbf{F})-\left|{z}_{k}\right|^{2}\sigma_{k}^{2},
	\end{equation}
	
	\vspace{-0.1cm}
	\noindent
	which satisfies $\tilde{\gamma}({z}_{k},\mathbf{W},\mathbf{F})\geq\underline{\phi}_{k}^{(i+1)}(\mathbf{F}),$ and is tight at $\mathbf{F}^{(i)}$ i.e. $\tilde{\gamma}({z}_{k},\mathbf{W},\mathbf{F}^{(i)})=\underline{\phi}_{k}^{(i+1)}(\mathbf{F}^{(i)})$. 
	Thus, in the $(i+1)$-th iteration, with the value of $\mathcal{Z}$ updated in closed form according to \eqref{Opt_eta}, problem (P4) can be approximated as
	
	\vspace{-0.5cm}	
		\begin{subequations} 
			\label{SubPro2_n:DeflctionAngOptCVX}
			\begin{align}
				\text{(P5):}~\underset{t,\mathbf{F}}{\text{max}}\quad&
				t \label{SubPro2_n:Obj} \\
				\text{s.t.} \quad&	
				\underline{\phi}_{k}^{(i+1)}(\mathbf{F})\geq t,\;\forall k \in \mathcal{K},\label{SubPro2_n:constr1}\\
				\phantom{\text{s.t.}\quad}&\|\mathbf{f}_{n}\|\leq1,\:\forall n \in \mathcal{N},\label{SubPro2_n:constr2}\\
				\phantom{\text{s.t.}\quad}&\eqref{SubPro2:constr2}.
			\end{align}
		\end{subequations}
	Note that constraint \eqref{SubPro2_n:constr2} is the relaxation of \eqref{SubPro2:constr3}, which aims to convert the original non-convex constraint into a convex one, and constraint \eqref{SubPro2_n:constr1} is a convex subset of \eqref{SubPro2:constr1}.

	Therefore, problem~(P5) is convex and can be solved via existing solvers such as CVX~\cite{grant2014cvx}. As a consequence of relaxing the equality constraint \eqref{SubPro2:constr3}, the optimum value of problem (P5) serves an upper bound of problem (P4). We normalize each pointing vector by $\mathbf{f}_{n}\leftarrow\frac{\mathbf{f}_{n}}{\|\mathbf{f}_{n}\|}$ in Algorithm~\ref{alg1} to ensure the obtained solution to be a feasible solution to problem (P1).
	
	\begin{algorithm}[!t]
		\caption{AO Algorithm for Solving (P1).}
		\begin{algorithmic}[1]\setAlgoSmall\footnotesize \label{alg1}
			\STATE \textbf{Initialization:} $\mathbf{F}^{(0)}=[\mathbf{e}_{x},\ldots,\mathbf{e}_{x}]$, random $\mathbf{W}^{(0)}$, threshold $\epsilon>0$, $i=0$.
			\STATE Compute $\mathcal{Z}^{(0)}=\{z_{k}^{\star}(\mathbf{W}^{(0)},\mathbf{F}^{(0)})\}_{k\in\mathcal{K}}$ via \eqref{Opt_eta}.
			\REPEAT
			\STATE Given $(\mathbf{F}^{(i)},\mathcal{Z}^{(i)})$, solve (P3) to obtain $\mathbf{W}^{(i+1)}$.
			\STATE Update $\mathcal{Z}^{(i)}\leftarrow\{z_{k}^{\star}(\mathbf{W}^{(i+1)},\mathbf{F}^{(i)})\}_{k\in\mathcal{K}}$.
			\STATE Given $(\mathbf{W}^{(i+1)},\mathcal{Z}^{(i)})$, solve (P5) to obtain $\mathbf{F}^{(i+1)}$.
			\STATE Update $\mathcal{Z}^{(i+1)}\leftarrow\{z_{k}^{\star}(\mathbf{W}^{(i+1)},\mathbf{F}^{(i+1)})\}_{k\in\mathcal{K}}$; $i\leftarrow i+1$.
			\UNTIL The fractional increase of the max-min SINR in \eqref{OptProOri:obj} falls below a threshold $\epsilon>0$ or the iteration number $i$ reaches the pre-designed number of iterations $I_{\max}$.
			\STATE Normalize $\mathbf{f}_{n}\leftarrow\mathbf{f}_{n}/\|\mathbf{f}_{n}\|$, $\forall n \in \mathcal{N}$.
			\STATE \textbf{Output:} $\mathbf{W}=\mathbf{W}^{(i)}$, $\mathbf{F}=\mathbf{F}^{(i)}$.
		\end{algorithmic}
	\end{algorithm}
	
	\vspace{-0.2cm}
	\subsection{Overall Algorithm}
	The overall AO algorithm to solve problem (P1) is summarized in Algorithm \ref{alg1}. Since this algorithm produces non-decreasing objective values of problem (P1) and the objective function is upper bounded. Therefore, Algorithm \ref{alg1} is assured to converge at a suboptimal solution. The complexity order of problem (P3) is $\mathcal{O}(\ln\frac{1}{\varepsilon}K^{1.5}M^{3}N^{3})$ , while that of problem (P5) is $\mathcal{O}(\ln\frac{1}{\varepsilon}KN^{3.5})$. Therefore, the overall complexity of solving (P1) is $\mathcal{O}(I\ln\frac{1}{\varepsilon}(KN^{3.5}+K^{1.5}M^{3}N^{3}))$, where $\varepsilon>0$ is a prescribed solution tolerance parameter for the interior-point method, and $I$ is the number of iterations required for convergence.
	
	\vspace{-0.2cm}
	\section{Simulation Results}
	\begin{figure}[h]
		\centering
		\subfloat[]{\includegraphics[width=0.24\textwidth]{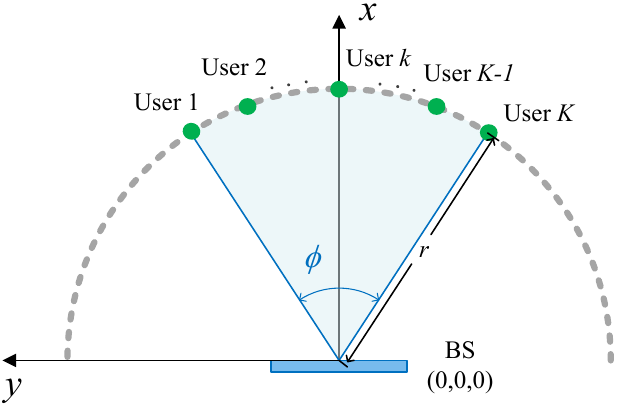}\label{fig2}}
		\hfil
		\subfloat[]{\includegraphics[width=0.23\textwidth]{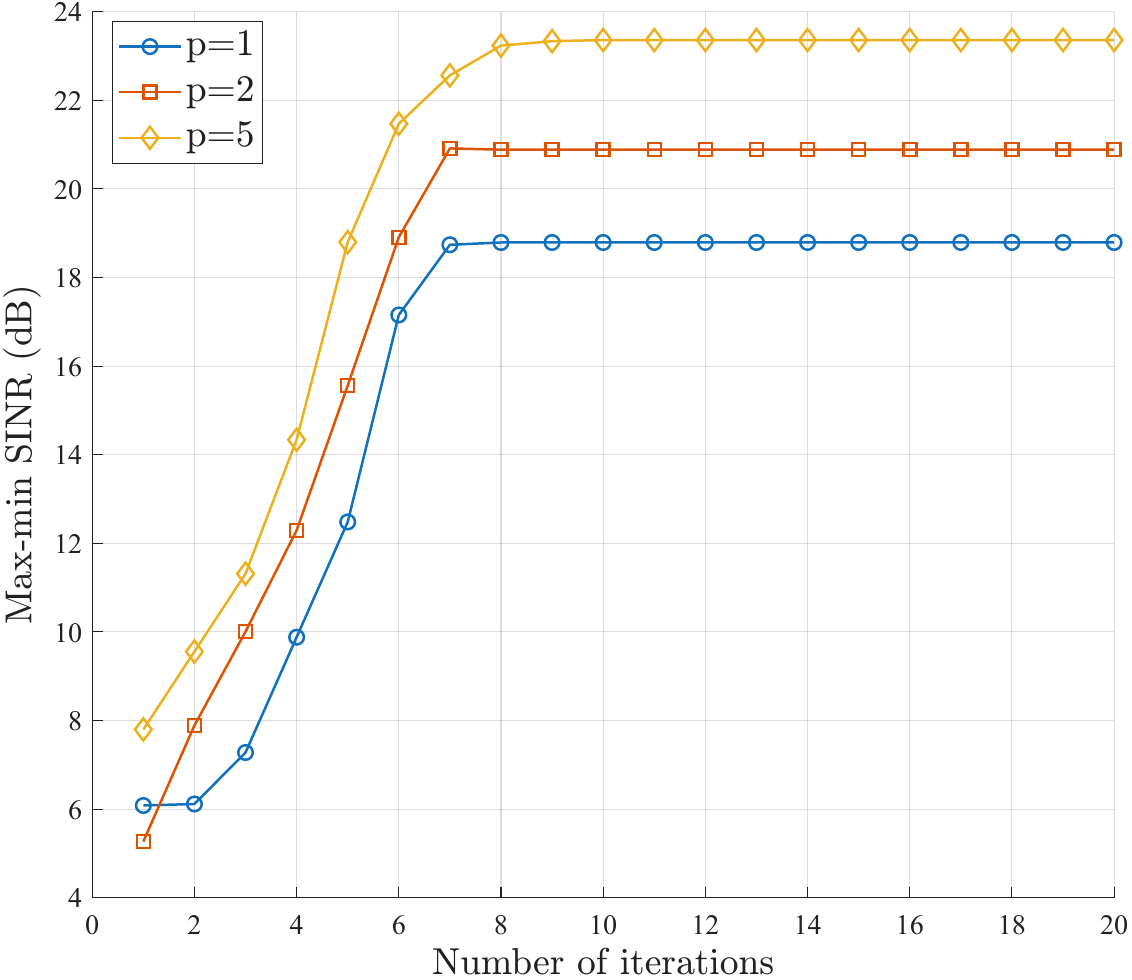}\label{fig_cov}}
		\hfil
		\caption{(a)~The top view of simulation setup. (b)~Convergence behavior of the proposed AO algorithm.}
		\label{tpview_fig_cov}
	\end{figure}
	In this section, we present simulation results to validate the effectiveness of our proposed RA-enhanced multicast wireless communication system as well as the proposed algorithm. We assume the system operates at a frequency of 2.4\:GHz with noise power $\sigma_{k}^{2}=-94$ dBm. The RA array is centered at the origin, with an inter-antenna spacing of $d=\frac{\lambda}{2}$. As shown in Fig.~\ref{tpview_fig_cov}~(a), users are uniformly distributed along a circular arc of radius $r>0$ on the plane $z = -h_0$, where $h_0 > 0$ denotes the height of the BS. This arc spans a user distribution angle $\phi \in (0,\pi]$ and is centered at $[0,0,-h_0]$. We set $h_0=10\:\si{\metre}$, $r=50\:\si{\metre}$ and $S=\frac{\lambda^2}{4\pi}$ as our global simulation parameter setting. Unless otherwise specified, we set $\phi=\frac{2\pi}{3}$, $P_{t}=15\:\text{dBm}$, the group size $|\mathcal{G}_{m}|=2,\forall m\in\mathcal{M}$, $M=2$, $N=4$, $p=5$, and $\theta_{\max}=\frac{\pi}{3}$.
	
	To evaluate the performance of the proposed RA-enhanced multicast communication system, referred to as \textbf{RA-based scheme}, we consider the following three benchmark schemes, all of which employ the optimal beamforming described in Section III-B: 1) \textbf{Fixed directional antenna-based scheme}: The orientations of all RAs are fixed, and $\mathbf{f}_{n}=\mathbf{e}_{x},\:\forall n \in\mathcal{N}$. 2) \textbf{Random orientation scheme}: The orientation of each RA is randomly generated within the rotational ranges specified in \eqref{OptProOri:constr1}, and simulation results are obtained by averaging over 100 independent random boresight realizations. 3) \textbf{Isotropic antenna-based scheme}: The directional gain is set to $G_0=1$ with $p=0$ in \eqref{GenericGainPattern}.
	
	In Fig.~\ref{tpview_fig_cov}\:(b), we plot the convergence behavior of the proposed AO algorithm under different directivity factors $p$. It can be observed that the SINR increases monotonically and converges within about 10 iterations. Moreover, the max-min SINR improves as $p$ increases. This is because a larger $p$ corresponds to a narrower beamwidth and a higher directional gain. By exploiting the rotational DoFs to adjust the boresight directions, the system can better balance the directional gains across users in the intended group while suppressing inter-group interference, thereby improving the max-min SINR.
	
\begin{figure*}[!t]
		\centering
		\subfloat[]{\includegraphics[width=0.31\textwidth]{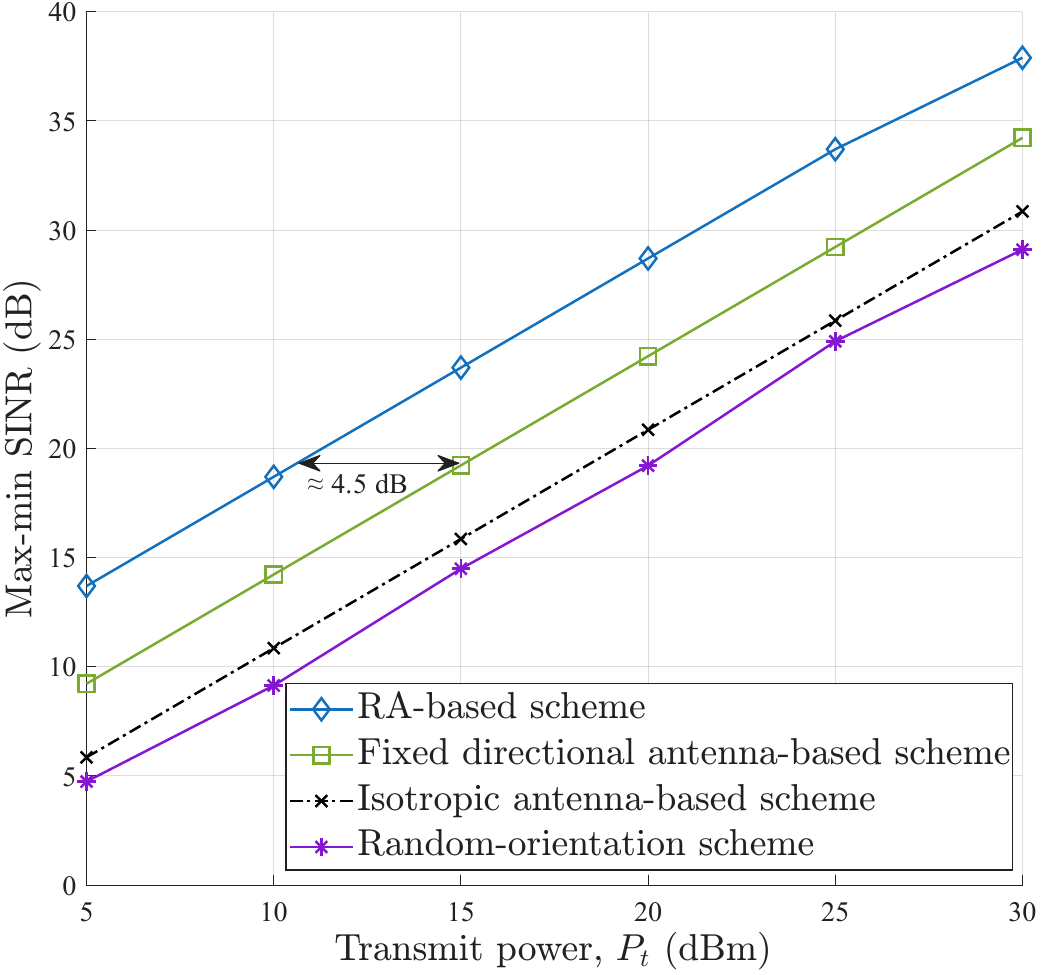}\label{P_t_SINR}}
		\hfil
		\subfloat[]{\includegraphics[width=0.32\textwidth]{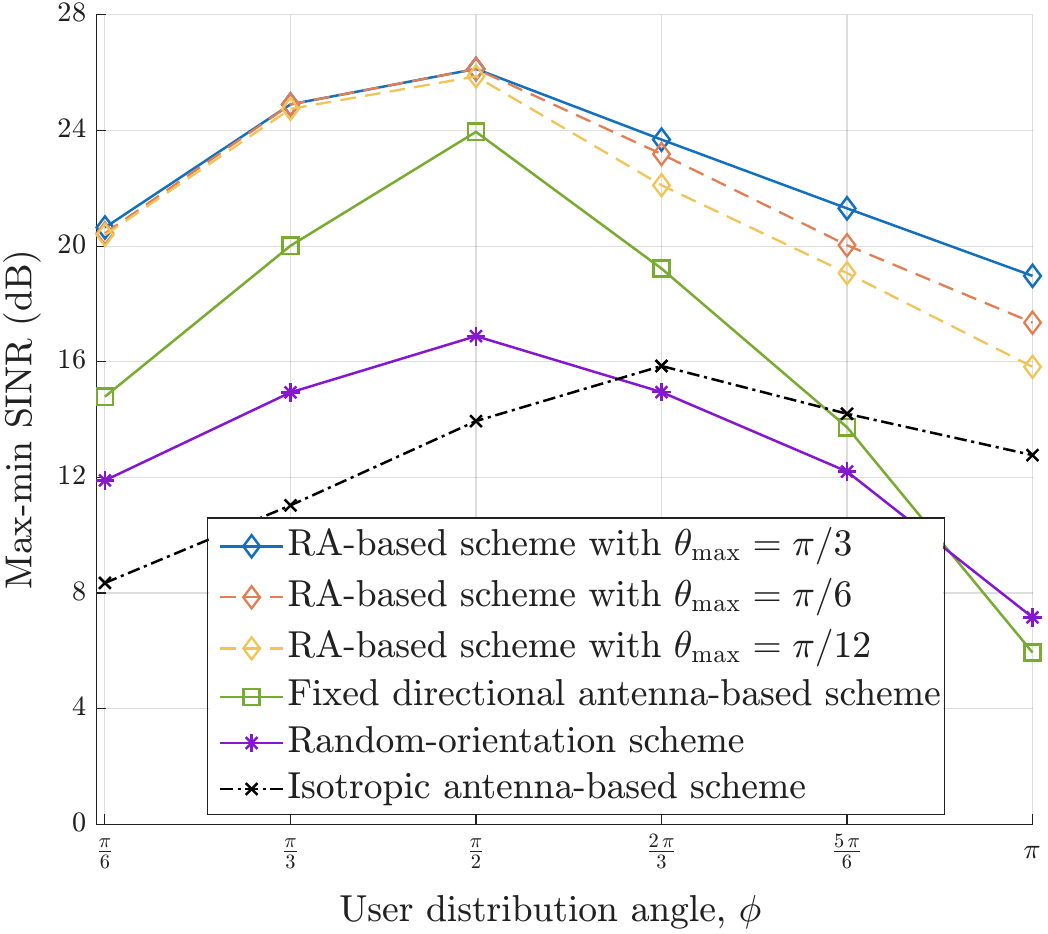}\label{User_dis}}
		\hfil
		\subfloat[]{\includegraphics[width=0.31\textwidth]{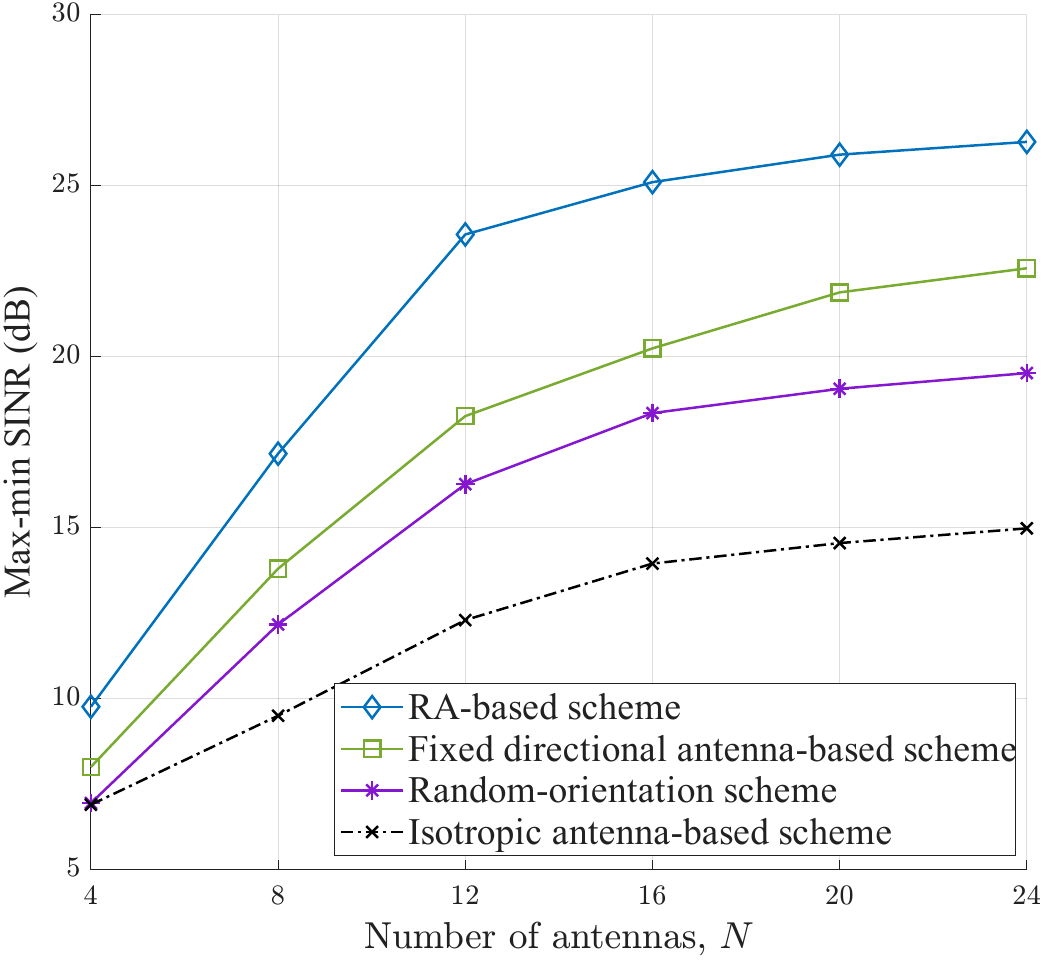}\label{NumAnt_SINR}}
		\vspace{-5pt}
		\caption{(a)~Average max-min SINR versus the maximum transmit power at the BS.
			(b)~Average max-min SINR versus the user distribution angle \texorpdfstring{$\phi$}{phi}.
			(c)~Average max-min SINR versus the number of antennas at the BS.}
		\label{fig:sim_results}
		\vspace{-12pt}
	\end{figure*}

	In Fig.~\ref{fig:sim_results}\:(a), we show the max-min SINR versus the maximum transmit power $P_{t}$. It is observed that as $P_{t}$ increases, the average max-min SINR of all four schemes rises monotonically. Notably, the proposed RA-based scheme outperforms the other three schemes over the entire $P_{t}$ range. This is expected, since RA introduces new DoFs in spatial orientation compared to the other schemes. With appropriate orientation optimization, its performance should be no worse than those methods without boresight adjustment. More specifically, the RA-based scheme achieves the same max-min SINR with approximately 4.5\:dB less power compared to its fixed-orientation counterpart. Furthermore, the random orientation scheme, which lacks proper optimization, exhibits the worst performance, which highlights the importance of optimizing the boresight direction.
	
	Fig.~\ref{fig:sim_results}\:(b) plots max-min SINR over the user distribution angle $\phi$. It is observed that all RA-based
	schemes still consistently outperform the other schemes. When $\phi < \frac{\pi}{2}$, all schemes exhibit improved performance as $\phi$ increase due to the reduced inter-group interference. As $\phi$ exceeds $\frac{\pi}{2}$, the performance of all RA-based schemes declines as the increased user separation complicates boresight alignment. Moreover, a larger $\theta_{\max}$ consistently yields higher max-min SINRs. In particular, the scheme with $\theta_{\max}=\frac{\pi}{3}$ is most robust against increasing $\phi$, underscoring the benefit of greater rotation range for each antenna. Even limited rotation ranges (e.g., $\theta_{\max}=\frac{\pi}{6}$ and $\frac{\pi}{12}$) provide meaningful gains over the fixed directional antenna-based scheme, which degrades sharply as $\phi$ increases. Conversely, the isotropic antenna-based scheme remains stable, surpassing other non-RA schemes at high angles (e.g., $\phi=\frac{5\pi}{6}$ and $\pi$). Overall, RA-based schemes demonstrate superior fairness-oriented performance across diverse user spatial distributions.
	
	In Fig.~\ref{fig:sim_results}\:(c), we present the average max–min SINR versus the number of BS antennas $N$. We set $M=3$, and $|\mathcal{G}_{1}|=|\mathcal{G}_{2}|=|\mathcal{G}_{3}|=4$. Across the whole range of $N$, the RA-based scheme still provides the highest SINR and exhibits the most rapid performance improvement when $N$ increases from $4$ to $12$. Due to the power constraint, all schemes experience gradually diminishing increment as $N$ increases. Notably, when $N=12$, the RA-based scheme is already superior to all other schemes across the range of $N$ shown in Fig.~\ref{fig:sim_results}\:(c). These results confirm the effectiveness of the proposed RA-scheme in improving the SINR performance under the given power constraint.
	
	\vspace{-0.3cm}
	\section{Conclusion}
	In this letter, we proposed an RA-enhanced multi-group multicast system, where the boresight direction of each antenna can be adjusted to flexibly steer the directional gain pattern. To maximize the minimum SINR among all the users, we develop an AO algorithm that jointly optimize the multicast beamforming and the boresight directions of RAs. Simulation results verified that by adjusting each RA's boresight direction, the proposed RA-based scheme can dynamically balance the directional gains among multicast users, thereby improving max-min fairness.
	
	\vspace{-0.4cm}
	{\appendix[Construction of \texorpdfstring{$L_{k}^{\mathrm{S}}$}{L\-\{k\}\^S} and \texorpdfstring{$L_{k,j}^{\mathrm{I}}$}{L\-\{k,j\}\^I}]
		\subsection{Construction of \texorpdfstring{$L_{k}^{\mathrm{S}}$}{L\-\{k\}\^S}}
		First, we derive the Hessian matrix $\nabla^{2}u_{k}(\mathbf{F})$ by calculate its diagonal blocks $\left\lbrace \nabla_{\mathbf{f}_{n}}^{2} u_k(\mathbf{F}) \right\rbrace_{n \in \mathcal{N}}$ as
		
		\vspace{-0.3cm}
		\begin{small}
			\begin{align}
				\abovedisplayshortskip=0pt
				\belowdisplayshortskip=0pt
				\abovedisplayskip=0pt
				\belowdisplayskip=0pt
				\nabla_{\mathbf{f}_n}^{2} u_k(\mathbf{F}) &= c_{k,n}\nabla^{2}g_{k,n}(\mathbf{f}_{n})\nonumber\\
				&=c_{k,n}p(p-1)\psi_{k,n}^{p-2}\vec{\mathbf{u}}_{k,n}\vec{\mathbf{u}}_{k,n}^{T},
			\end{align}
		\end{small}\normalsize
		
		\vspace{-0.2cm}
		\noindent
		where $\psi_{k,n}\triangleq\mathbf{f}_{n}^{T}\vec{\mathbf{u}}_{k,n}$. Since the off-diagonal blocks vanish i.e., $\nabla_{\mathbf {f}_{\tilde{n}}}\nabla_{\mathbf f_n}^{\!T} s_k(\mathbf F)=\mathbf 0_{3\times3},\:\tilde{n}\neq n,\forall \tilde{n},\:n \in \mathcal{N}$, $\nabla^{2}u_{k}(\mathbf{F})$ is block diagonal, which is given by
		
		\vspace{-0.3cm}
		{\footnotesize
			\begin{equation}
				\nabla^{2}u_{k}(\mathbf{F}) = \text{blkdiag}\left(\nabla_{\mathbf{f}_1}^{2} u_k(\mathbf{F}),\nabla_{\mathbf{f}_2}^{2} u_k(\mathbf{F}),\ldots,\nabla_{\mathbf{f}_N}^{2} u_k(\mathbf{F})\right).
			\end{equation}
		}\normalsize
		
		\vspace{-0.1cm}
		\noindent
		The Euclidean norm of $\nabla^{2}u_{k}(\mathbf{F})$ can be upper bounded as
		
		\vspace{-0.5cm}
		\begin{small}
			\begin{align}
				\label{ieq:L2HessU_k}
				\lVert\nabla^{2}u_{k}(\mathbf{F})\rVert&= \underset{n}{\max}\;\|\nabla_{\mathbf{f}_{n}}^{2}u_{k}(\mathbf{F})\|\leq C_{\text{max},k}p\lvert p-1\rvert\triangleq\bar{L}_{k}^{\mathrm{S}},
			\end{align}
		\end{small}
		
		\vspace{-0.2cm}
		\normalsize
		\noindent
		where $C_{\text{max},k}=\underset{n}{\max}\;\lvert c_{k,n}\psi_{k,n}^{p-2} \rvert$. According to \cite{sunMajorizationMinimizationAlgorithmsSignal2017a}, we select $ L_{k}^{\mathrm{S}}$ to ensure $L_{k}^{\mathrm{S}}\:\mathbf{I}_{3}\succeq\nabla^{2}u_{k}(\mathbf{F})$, thereby satisfying inequality \eqref{ieq:lb_uk}. Since $\bar{L}_{k}^{\mathrm{S}}\mathbf{I}_{3}\succeq\lVert\nabla^{2}u_{k}(\mathbf{F})\rVert\mathbf{I}_{3}\succeq\nabla^{2}u_{k}(\mathbf{F})$, we can simply choose $L_{k}^{\mathrm{S}}=\bar{L}_{k}^{\mathrm{S}}$.
		
		\vspace{-0.5cm}
		\subsection{Construction of \texorpdfstring{$L_{k,j}^{\mathrm{I}}$}{L\-\{k,j\}\^I}}
		
First, the Hessian matrix $\nabla^{2}a_{k}(\mathbf{F}) \in \mathbb{R}^{3N \times 3N}$ is structured as a block matrix:

\vspace{-0.3cm}
\begin{equation}
	\nabla^{2}a_{k}(\mathbf{F}) = \Big[ \mathbf{H}_{\tilde{n},n} \Big]_{1 \leq \tilde{n}, n \leq N}~,
\end{equation}

\vspace{-0.2cm}
\noindent
where the $(\tilde{n},n)$-th block $\mathbf{H}_{\tilde{n},n}$ is computed as
		
		\vspace{-0.3cm}
		{\footnotesize
			\begin{align}\label{eq:H_blocks}
				\mathbf{H}_{\tilde{n},n}
				=
				\begin{cases}
					\Big[\,2\operatorname{Re}\{x_{k,j}^{*}\upsilon_{k,j,n}\}\,p(p-1)\psi_{k,n}^{p-2}\\
					+2\,|\upsilon_{k,j,n}|^{2}p^{2}\psi_{k,n}^{2(p-1)}\Big]\vec{\mathbf{u}}_{k,n}\vec{\mathbf{u}}_{k,n}^{T},
					& \tilde{n}=n,\\[0.5em]
					2\operatorname{Re}\!\{\upsilon_{k,j,n}\upsilon_{k,j,\tilde{n}}^{*}\}\,p^{2}
					\psi_{k,n}^{p-1}\psi_{k,\tilde{n}}^{p-1}\;
					\vec{\mathbf{u}}_{k,n}\vec{\mathbf{u}}_{k,\tilde{n}}^{T},
					& \tilde{n}\neq n,
				\end{cases}
		\end{align}}\normalsize where $\upsilon_{k,j,n}\triangleq \tilde{\beta}_{k,n}w_{j,n}$. 
		Similar to \eqref{ieq:L2HessU_k}, 
	the norm of the Hessian matrix $\nabla^2 a_{k,j}(\mathbf{F})$ can be upper bounded as
	\vspace{-0.2cm}
	{\footnotesize
		\begin{align}\label{eq:Lg}
			\left\| \nabla^2 a_{k,j}(\mathbf{F}) \right\| &\le \underset{{\tilde{n}}}{\text{max }}\,\sum\limits_{n=1}^{N}{\left\| {\mathbf{H}_{\tilde{n},n}} \right\|} \nonumber \\[-8pt]
			&\leq 2p(|p-1|+p)V_{\max,k,j}\sum_{i=1}^{N}{\left| {{\upsilon}_{k,j,i}} \right|}\triangleq {\bar{L}^{\mathrm{I}}_{k,j}},
	\end{align}}\normalsize 

\vspace{-0.2cm}
\noindent
where $V_{\max,k,j}=\underset{n}{\max \,}|v_{k,j,n}|\psi_{k,n}^{2(p-1)}$. Therefore, $\bar{L}_{k,j}^{\mathrm{I}}\mathbf{I}_{3}\succeq\lVert\nabla^2 a_{k,j}(\mathbf{F})\rVert\mathbf{I}_{3}\succeq\nabla^2 a_{k,j}(\mathbf{F})$ holds, then ${L}_{k,j}^{\mathrm{I}}=\bar{L}^{\mathrm{I}}_{k,j}$ can be chosen~\cite{sunMajorizationMinimizationAlgorithmsSignal2017a}.

}

\vspace{-0.4cm}
\bibliographystyle{IEEEtran}
\bibliography{Refs/Refs.bib}

\end{document}